# Light fan driven by relativistic laser pulse

Yin Shi, Baifei Shen*, Lingang Zhang,  Xiaomei Zhang, Wenpeng Wang, and Zhizhan Xu

*State Key Laboratory of High Field Laser Physics, Shanghai Insitute of Optics and Fine Mechanics, Chinese Academy of Sciences, Shanghai 201800, China*

**Abstract**

When a relativistic laser pulse with high photon density interacts with a specially tailored thin foil target, a strong torque is exerted on the resulting spiral-shaped foil plasma, or "light fan". Because of its structure, the latter can gain significant orbital angular momentum (OAM), and the opposite OAM is imparted to the reflected light, creating a twisted relativistic light pulse. Such an interaction scenario is demonstrated by particle-in-cell simulation as well as analytical modeling, and should be easily verifiable in the laboratory. As important characters, twisted relativistic light pulse has strong torque and ultra-high OAM density.

Prompted by the fast development of laser techniques [1], light-matter interaction has entered the regime of relativistic laser-plasma interaction. Over the past few decades, a number of novel mechanisms and schemes have been proposed. Among these mechanisms and schemes, the most promising application is for use in laser-driven plasma accelerator science, such as laser wakefield acceleration (LWFA) of electrons [2] and laser driving foil to accelerate protons [3]. Laser-plasma interaction can also be efficient sources of high-order harmonic generation (HHG) [4], x-rays [5], and even gamma-rays [6, 7]. One of the key issues in the above mechanisms is how to make use of the laser ponderomotive force efficiently to pump strong charge separation field in plasma, which is the origin of particle acceleration. Hence, it is the force (the accelerating force, the confining force etc.) that people care most in relativistic laser plasma physics. The effect of another important dynamical quantity, the torque, although as important as force, has not been revealed for relativistic laser pulse. How to observe the orbital angular momentum (OAM) in laser-plasma interaction and how the appearance of OAM would essentially affect the process are of special interests. Circularly polarized light carries a spin angular momentum of $\pm\hbar$ per photon; however, the total OAM of a normal Gaussian pulse, commonly found in the current chirped pulse amplification technology, is zero. Therefore, observation of the torque and OAM in relativistic laser-plasma interaction is rare.

OAM has been discussed extensively for weak light [8-13]. Since Allen et al. first showed that a Laguerre-Gaussian (LG) laser pulse has finite OAM [8], many applications using twisted lights have been found [9-11]. The OAM of a twisted light can be transferred to matter. More interestingly, several phenomena observed in astrophysics, like pulsars, are related to the OAM of light and plasma [14, 15]. Thus, simulating and investigating such immense process in a laboratory on earth would be of great convenience. Recently, Mendonca et al. have derived the solutions of plasma wave with OAM [16, 17]. Also electron beam with OAM has drawn many interests[18, 19].

Although several methods are available for obtaining twisted laser pulses, relativistic twisted light pulse has not been generated yet. The OAM effect of relativistic laser-plasma interaction remains unknown. In this letter, we propose a simple and effective method for generating a relativistic twisted laser pulse and showing the OAM effect. It is called "Light Fan": a relativistic laser pulse (with very high photon density) impinges on a spiral foil (the fan); hence, both the fan and the reflected pulse achieve a net large OAM. This study demonstrates, for the first time, that a reflect fan structure can be applicable in the relativistic regime. More importantly, the dynamic process of such structure presents new and unique features in the relativistic intensity regime.

The proposed scheme is verified by the following three-dimensional (3D) particle-in-cell (PIC) simulations. The set-up is sketched in Fig. 1. The foil consists of eight parts with the same step height $\Delta h = \lambda/16$ to mimic a $\lambda/2$ spiral phase plate. A relativistic laser pulse normally incidents on the foil from the left, interacts with the matter and is then reflected. The 0.8 μm laser pulse with Gaussian time envelope $a(t) = a_0 \exp(-t^2/\Delta t^2)$ is linearly polarized along the y axis, and has a peak amplitude of $a_0 = 5$, pulse duration of $\Delta t = 13.3$ fs, and spot size of $w = 3$ μm, where $a_0 = eE_y/m_e\omega_0 c$ is the normalized dimensionless laser electric field. The spiral foil has a density of $100 n_c$, where $n_c = m_e\omega_0^2/4\pi e^2$ is the critical density. The simulation box is 15μm×20μm×20μm in the $x \times y \times z$ directions, respectively. The foil is initially located between $x = 7.0$ μm and $x = 7.8$ μm. The simulation mesh size is d$x$ = 0.025 μm in the x direction and d$y$ = d$z$ = 0.05 μm in the $y$ and $z$ directions. The number of macro particle per cell is PPC = 2.

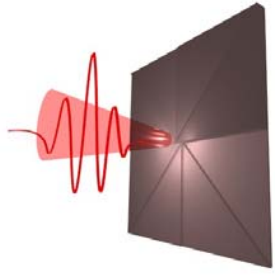

Fig. 1. The foil used in the simulation has eight parts, with the same step height $\Delta h = \lambda/16$ to mimic a $\lambda/2$ spiral phase plate.

First, the distributions of $E_y$ in $y$-$z$ plane of $x = 3.0$ μm (Fig. 2(a)) and $x$-$z$ plane of $z = 0.0$ μm (Fig. 2(b)) at time $t = 51.3$ fs when beam is totally reflected are provided. It shows clearly the characters of twisted light. When the plasma surface is seen as a structured mirror, which is an approximate assumption in many cases, the reflected light mode can be expanded with a series of LG modes [20]. The amplitude $LG_{nm}$ is defined by

$$\begin{aligned}\mathrm{LG}_{nm}(\rho,\phi,x) = (C_{nm}/w)\exp(-ik\rho^2/2R)\exp(-\rho^2/w^2)\exp[-i(n+m+1)\psi] \times \\ \exp[-i(n-m)\phi](-1)^{\min(n,m)}(\rho\sqrt{2}/w)^{|n-m|}L_{\min(n,m)}^{|n-m|}(2\rho^2/w^2),\end{aligned} \quad (1)$$

with $R(x) = (x_R^2 + x^2)/x$, $\frac{1}{2}kw^2(x) = (x_R^2 + x^2)/x_R$, $\psi(x) = \arctan(x/x_R)$, where $C_{nm}$ is a normalization constant, $k = 2\pi/\lambda$ the wave number, $x_R$ the Rayleigh range, and $L_p^l(x)$ is the generalized Laguerre polynomial. The Gaussian mode of incident laser beam in simulation is $LG_{00}$ mode. The mode decomposition of a $LG_{00}$ mode whose wavefront has been modified by the fan structure is then defined by the expansion coefficients [20]

$$\begin{aligned}a_{st} &= \langle \mathrm{LG}_{st} | \exp(-i\Delta\phi) | \mathrm{LG}_{00} \rangle \\ &= \iint \rho\,d\rho\,d\phi (C_{st}^*/w_{st})\exp(ik\rho^2/2R_{st} - \rho^2/w_{st}^2)\exp[-i(s-t)\phi] \times \\ &\quad (-1)^{\min(s,t)}(\rho\sqrt{2}/w_{st})^{|s-t|}L_{\min(s,t)}^{|s-t|}(2\rho^2/w_{st}^2)\exp(-i\Delta\phi) \times \\ &\quad (C_{00}/w)\exp(-ik\rho^2/2R - \rho^2/w^2),\end{aligned} \quad (2)$$

where

$$\Delta\phi = \sum_{n=0}^{7} H(\phi - \frac{\pi}{4}n) H(\frac{\pi}{4}n + \frac{\pi}{4} - \phi)\frac{2\pi}{8}n. \quad (3)$$

Because of the fact that $\Delta\phi \approx \phi$, only the modes with $l = s - t = 1$ will contribute most in the $\phi$ integral. According to the animation (Supplementary Information, distributions of $E_y$ in $y$-$z$ plane changing with different $x$ ($x = i \cdot dx$) and laser wavelength corresponding to $\Delta i = \lambda/dx = 32$), the number of intertwined helices can be found to be $l = 1$. In the numerical calculation of Eq. (2), the Rayleigh range and the waist of the reflected light are assumed to equal to those of the incident beam, respectively (i.e., $R_{st} = R$, $w_{st} = w$). The relative weight of the modes is given by $I_{st} = |a_{st}|^2$. Our calculations show that $I_{10} \approx 64.8\%$ and $I_{21} \approx 13.6\%$. We obtain a good approximation on the distributions of the reflected light electric field using $\sqrt{0.65}$ LG$_{10}$ + $e^{i\theta}\sqrt{0.14}$ LG$_{21}$, where $\theta$ is the relative phase chosen to make the approximation closer to the simulation results (in our case, $\theta = \pi/2$).

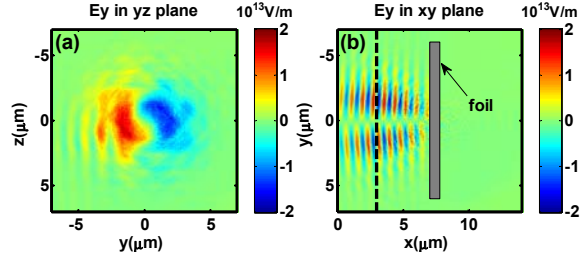

Fig. 2 (color online). Electric field $E_y$ distribution in (a) $y$-$z$ plane of $x = 3.0$ μm and (b) $x$-$y$ plane of $z = 0.0$ μm at $t = 51.3$ fs when beam is totally reflected. The black dashed line shows the $x$ position of the plane in (a). An animation in the Supplementary Information is provided by changing the $x$ ($x = i \cdot dx$) position of plane in (a).

In the theory of LG beam, the ratio of the angular momentum flux $J$ to the energy flux $P_L$ is $J/P_L = l/\omega$ [8]. According to the laser parameters in the simulations, the total OAM is about $L = lP_L\Delta t/\omega = 3.0 \times 10^{-17}$ kg m$^2$ s$^{-1}$, which agrees well with the OAM sum of electrons and protons in scale and direction (Fig. 3). In addition, the total OAM of the reflected light is calculated based on the electromagnetic field data from the PIC simulations. The result is $\vec{L_x} = \varepsilon_0 \iint \vec{r} \times (\vec{E} \times \vec{B}) dt d\vec{r} = 1.1 \times 10^{-17}$ kg m$^2$ s$^{-1}$, which is close to the result of LG beam theory.

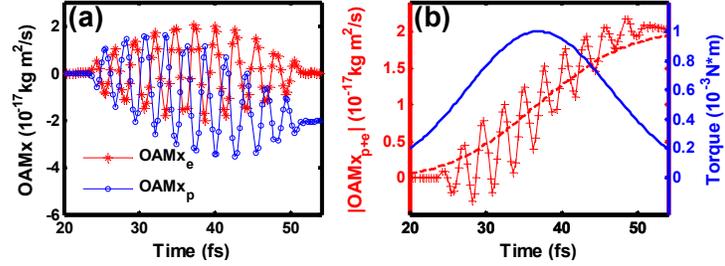

Fig. 3 (color online). (a) OAM (in the $x$ direction) of electrons (red stars) and protons (blue circles) changing with time. (b) Sum of the OAMs of electrons and protons (red pluses) changing with time; the blue solid line is fitted based on the assumption that torque changes with longitudinal laser pulse envelope $\tau_L = 10^{-3} \exp\{-[(t - 37\text{fs})/13.3\text{fs}]^2\}$ N m . Integration of torque gives the fitting OAM curve (red dashed).

The statistical OAM in the $x$ direction of the electrons and protons $L_x = \sum (\vec{r} \times \vec{p})_x$ at different times is given in Fig. 3(a). It shows that the OAM of both electrons and protons are oscilating with time, and the frequency is the same as driven laser frequency. However, the proton OAM falls behind a phase $\pi$ relative to electron OAM. The former obtains a net OAM $L_{px} = -2.0 \times 10^{-17}$ kg m$^2$ s$^{-1}$ (blue circles in Fig. 3(a)) and the latter only osciates around zero (red stars in Fig. 3(a)). On the surface of the foil, any two opposite parts that are symmetric about the centre will feel the laser field with $\pi$ phase difference, resulting in a significant OAM. As the laser field wave interacts with the foil, both electrons and protons rotate clockwise and anti-clockwise under the action of the laser field and charge seperation field. As we know, the angular momentum of a rotating disk is $L = 0.5 r^4 \rho h \omega$, where $r$ is the radius of disk, $\rho$ the density, $h$ the thickness, and $\omega$ is the angular velocity. If we see the rotator as a protons disk with $r$ = 3 μm , $h$ = 0.8 μm and $\rho$ = 100$m_p n_c$, the angular velocity corresponding to $L = 2.0 \times 10^{-17}$ kg m$^2$ s$^{-1}$ is about $\omega = 7 \times 10^8$ rad s$^{-1}$. It can be one order higher than man-made rotational speed record $6 \times 10^7$ rad s$^{-1}$, which is also laser-induced [21]. The rotation of the plasma after the laser pulse leaves may be used to simulate the pulsars [15].

Fig. 3(b) (red pluses) shows the sum of the OAMs of electrons and protons $L_x = -2.0 \times 10^{-17}$ kg m$^2$ s$^{-1}$, which is same with the calculated OAM of reflected beam in scale and is opposite in direction. The fitting curve of the total OAM (red dashed line in Fig. 3(b)) is based on the assumption that torque $\tau_L = \partial L_x / \partial t$ changes with longitudinal laser pulse envelope $\tau_L = 10^{-3} \exp\{-[(t - 37\text{fs})/13.3\text{fs}]^2\}$ N m (blue solid line in Fig. 3(b)), where 37 fs is the arrival time of the laser peak on the foil surface. The peak torque corresponds to the peak intensity of the laser, and the fastest variation of OAM. The head and tail of torque distribution corresponds to the head and tail of laser pulse envelope, and the flat part of OAM. A torque of $10^{-3}$ N m can be given to a disk with a radius about 3 μm by such a relativistic laser. It indicates that a huge OAM can be transported to a small volume in very short time. The twisted light with strongest torque by far can be found in twisted laser driven HHG experiments. In the experiments of Zurch et al. [21], a 800 nm laser is reflected from the spatial light modulator to obtain a twisted light with peaking intensities of approximately $2 \times 10^{15}$ W cm$^{-2}$ focusing on a spot size of 40 μm. Assuming the best transformation (Gaussian mode totally converted to a series of LG modes with $l = 1$), the torque that such intensity can provide to a 3 μm disk is about $8 \times 10^{-8}$ N m. So the torque of relativistic twisted light in our plan can be 4-orders higher. In our case, the OAM density of relativistic twisted light, defined with OAM over volume, can be as high as $u = L_x / V = 0.56$ kg m$^{-1}$ s$^{-1}$. Considering that high energy density is an important concept, this high OAM density may bring some new effects. Similar to the ponderomotive force originating from inhomogeneity of laser intensity, torque comes from both the laser intensity distribution and helical phase fronts. Torque is an intrinsic characteristic like the ponderomotive force. Thus, we should expect that the torque characteristic will be as important as ponderomotive force in relativistic light-matter interaction, and brings out some potential applications.

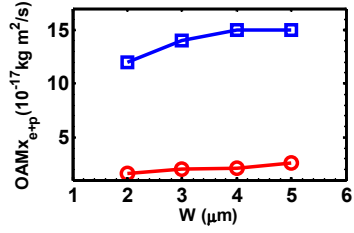

Fig. 4 (color online). The effect of laser spot size. The results of the two different powers, $P_0 = 4.8 \times 10^{12}$ W (blue squares) and $P_1 = 4.3 \times 10^{13}$ W (red circles), corresponding to $a_0 = 5$ and $a_1 = 15$, respectively, when the spot size is 3 μm.

Fig. 4 shows the changes of the total OAM when keeping the power unchanged for different spot sizes. The results of the two different powers, $P_0 = 4.8 \times 10^{12}$ W (blue squares) and $P_1 = 4.3 \times 10^{13}$ W (red circles), are shown. The total OAM almost also keeps unchanged for a constant power. It can be easily understood from the view of OAM conservation between light and plasma. For the light, the ratio of the angular momentum flux to the energy flux is only decided by the number of intertwined helices [8], then the total step height. So the same power brings the same OAM, and the higher power brings the higher OAM. The little decrease of OAM when spot size reduces can be understood as following. With smaller spot size and higher intensity, the structure of foil may deform more heavily in the later process of laser-plasma interaction. Given the deviation from the spiral step structure, the total OAM of matter or light decreases slightly. Based on theory of LG, the peak OAM density of LG beam is $u = lI_0/(c\omega)$, where $I_0$ is the peak laser intensity. Then the OAM density of relativistic twisted beam can be very huge due to high intensity of relativistic beam. It also shows that the OAM density can be higher when laser beam focuses smaller.

Up to now, OAM is only emphasized in conventional optics. Apart from the phase change carried out in optics, this work reveals the dynamic process of relativistic laser interaction with plasma, which have completely not been elucidated. Combined with other schemes of relativistic mirrors in plasma (ion acceleration, coherent radiation with high energy, hot electron transport) [22], OAM in relativistic light-matter interaction may bring new effects. For example, relativistic attosecond X-ray vortices from HHG should also be expected. Consequential explorations about the effects of OAM on various aspects, such as transport of hot electrons, proton acceleration, and plasma wave, may become available. Using relativistic table-top lasers, this scheme creates a new laboratory plasma regime with extreme plasma parameters, such as huge OAM. It may be used to mimic some astrophysical environment such as the pulsars [15].

As an important character, relativistic laser has strong torque. This strong torque can lead to strong OAM when the laser interacts with structured matter (like Light Fan in our case) or when twisted light (like the reflected light in our case) interact with uniform matter. Like strong ponderomotive force of relativistic laser can produce a finite linear momentum in many applications (for example LWFA), such a strong torque should also be expected to produce a finite OAM in many potential applications. Also a relativistic twisted laser has an ultra-high OAM density accompanying ultra-high energy density. We expect experiments based on light-fan scheme can be realized soon, which should lead to some novel results on strong laser field physics [23].

This work was supported by the Ministry of Science and Technology (Grant Nos. 2011DFA11300, 2011CB808104), the National Natural Science Foundation of China (Grant Nos. 61221064, 11125526, 11335013, 11127901).

*To whom all correspondence should be addressed: bfshen@mail.shcnc.ac.cn